\documentclass[11pt,english]{article}

\usepackage[T1]{fontenc}
\usepackage[latin1]{inputenc}
\usepackage[english]{babel}
\usepackage{lmodern}
\usepackage{a4wide}
\usepackage{amssymb, amsmath, amsthm}
\usepackage{slashed}
\usepackage{float}
\usepackage{graphicx}
\usepackage{psfrag}
\usepackage{lscape}
\usepackage[all]{xy}
\usepackage{hyperref}
\usepackage{enumerate}
\usepackage{dsfont}
\usepackage{cite}
\usepackage{color}
\usepackage{upgreek}
\usepackage{datetime}

\newcommand{\beq}{\begin{equation}}
\newcommand{\eeq}{\end{equation}}
\def\bea#1\eea{\begin{align}#1\end{align}}

\makeatletter
\@addtoreset{equation}{section}
\makeatother

\begin{document}


\thispagestyle{empty}
\rightline{\small LMU-ASC 35/14}
\rightline{\small MPP-2014-222}

\vskip 2.1cm

\begin{center}

{\fontsize{18.2}{21}\selectfont{ \bf{On Inflation and de Sitter 
      \vskip0.3cm  in
      Non-Geometric String Backgrounds}}}

\vskip 2.1cm

\noindent{\textbf{  Falk Hassler$^a$, Dieter
    L\"ust$^{a,b}$ and Stefano Massai$^a$}}

\vskip 0.6cm
\noindent
\textit{$^a$ Arnold-Sommerfeld-Center for Theoretical Physics\\Fakult\"at f\"ur Physik, Ludwig-Maximilians-Universit\"at M\"unchen\\Theresienstra\ss e 37, 80333 M\"unchen, Germany}\\
\vskip 0.3cm\noindent
 \textit{$^b$ Max-Planck-Institut f\"ur Physik\\F\"ohringer Ring 6, 80805 M\"unchen, Germany}

\vskip 0.3cm
%
\noindent {\small{\texttt{f.hassler@lmu.de, dieter.luest@lmu.de, stefano.massai@lmu.de}}}

\vskip 2.5cm

\end{center}

\begin{abstract}

\noindent 
We study the problem of obtaining de Sitter and inflationary vacua from dimensional reduction of double field theory (DFT) on nongeometric string backgrounds. In this context, we consider a new class of effective potentials that admit Minkowski and de Sitter minima. We then construct a simple model of chaotic inflation arising from T-fold backgrounds and we discuss the possibility of trans-Planckian field range from nongeometric monodromies as well as the conditions required to get slow roll.

\vfill
\end{abstract}
\newpage

\tableofcontents

\section{Introduction}

It is clearly an important  task to derive  effective four-dimensional potentials from string theory that 
can describe all observed phenomenological properties of particle physics and cosmology. The list of phenomenological restrictions on the effective string potentials is rather long and challenging: one has to
break supersymmetry in a controllable way, to explain the Higgs mechanism and the Standard Model symmetry  breaking with great detail, as well as to address cosmological observations.

Focusing on applications to cosmology, we can list three main tasks: moduli stabilization, the creation of an inflationary potential and finally the generation of a small and positive cosmological constant. When discussing these problems, it is desirable to focus on the universal features rather than building on very model dependent details of particulars string vacua. Specifically, it is important to understand if the generic structure of the string landscape puts more severe restrictions on the form of the effective potentials compared to what comes out from gravity or supergravity in higher dimensions (for some discussion on this general issue see~\cite{Vafa:2005ui,Ooguri:2006in,ArkaniHamed:2006dz}). An important example is the difficulty in building a string model with trans-Planckian range for the inflaton field~\cite{Banks:2003sx,ArkaniHamed:2006dz,Baumann:2006cd}. In light of the recent cosmological observation of large tensor perturbation in the CMB reported by the BICEP2 collaboration, if confirmed, this restriction become crucial. Indeed, it is well known~\cite{Lyth:1996im} that an observation of a large tensor perturbation would imply a trans-Planckian excursion for the inflaton. It is then critical to understand possible ways to evade constraints on inflaton field range in the string landscape\footnote{See for example~\cite{Burgess:2013sla,Linde:2014nna,Baumann:2014nda,Kallosh:2014xwa} for reviews on (string) inflation.}.

In this paper we will consider a class of effective potentials which originate from internal fluxes in the universal NS sector of the ten-dimensional type II superstrings. In addition to $H$-fluxes, at the level of the effective theory one can add additional fluxes, such as $f$-fluxes (derivatives of non-constant components in the internal metric) and more exotic non-geometric $Q$-fluxes, which can be viewed as derivatives of a bi-vector $\beta$. These kind of backgrounds can be described as T-folds, or compactifications with duality twists~\cite{Dabholkar:2002sy,Hull:2006va}, which are in the simplest cases just fibrations of a two-dimensional torus $T^2$ over a one-dimensional base $S^1$. They are characterized by the monodromy of the $T^2$-fibre when going around the $S^1$-base. The monodromy is encoded in a twist matrix $M$, which determines how the T-duality group $O(2,2,\mathbb{Z})\sim SL(2,\mathbb{Z})\times SL(2,\mathbb{Z})$ acts on the fibre when encircling the base. We will argue that such non-trivial monodromies allow the corresponding moduli (the complex structure $\tau$ and the K\"ahler parameter $\rho$ of the fibre torus) in the effective scalar potential to take values in bigger domains than the fundamental domain of $SL(2,\mathbb{Z})_\tau\times SL(2,\mathbb{Z})_\rho$. The role of such monodromies, as first noticed in~\cite{Silverstein:2008sg,McAllister:2008hb}, has important consequences for inflation, since in principle it can allows trans-Planckian field range of the inflaton (which in~\cite{Silverstein:2008sg,McAllister:2008hb} is identified with an axion). We note that T-duality is responsible for the non-trivial monodromies that extend the field range, while often it is precisely T-duality that makes attaining large inflaton range problematic~\cite{Banks:2003sx}.

The background monodromies can belong to parabolic or elliptic conjugacy classes of the duality group~\cite{Dabholkar:2002sy,Schulgin:2008fv} (in addition there are also hyperbolic and sporadic classes).
 In the parabolic case, the monodromy has infinte order, 
and this situation allows for infinitely large field configurations. For example, the Nilmanifold used in the original axion monodromy model~\cite{Silverstein:2008sg} is a geometric space with infinitely large monodromy. Applied to wrapped D4 branes this is what extends the inflaton range. The effective NS potentials from constant fluxes with parabolic monodromies have the following schematic form (in the Einstein frame):
 \begin{equation}
V(\rho_I,\sigma) = \sigma^{-2} \left(\rho_I^{-3}\ V_H + \rho_I^{-1}\ V_f + \rho_I\ V_Q + \rho_I^{3}\ V_R  \right)\, . \label{eq:pot}
\end{equation}
 Here $\rho_I$ denotes then internal volume modulus, $\sigma$ corresponds to the dilaton field, and $V_H$, $V_f$, $V_Q$ and $V_R$ are the potential energies created by the corresponding constant fluxes $H$, $f$, $Q$, $R$ (being quadratic in the fluxes, i.e. $V_H\sim H^2$, \dots). Note that in general not all fluxes can be turned on at the same time, but only one kind of flux on a three-dimensional space can be non-vanishing. Simple no-go theorems~\cite{Hertzberg:2007wc,Caviezel:2008tf,Danielsson:2009ff} show that these potentials do not have the desired properties for inflation, and up-lifts  to stable Minkowski or de Sitter vacua are not possible.\footnote{This is related to the fact that the NS backgrounds with only parabolic fluxes do not solve the full string field equations, i.e. the $\beta$-function equations are only satisfied in lowest order of the fluxes.} However some ways to relax these no-go theorems via geometric as well as non-geometric fluxes were discussed in \cite{Haque:2008jz,Danielsson:2009ff,Danielsson:2011au,Blaback:2013ht,Damian:2013dq,Damian:2013dwa}. In addition, other ingredients such as NS 5-branes, anti NS 5-branes and KK monopoles can be introduced to stabilize some of the moduli and provide an uplift to a de Sitter vacuum~\cite{Silverstein:2007ac}.\footnote{More recently string inflation with D7-branes and in F-theory  was discussed in \cite{Hebecker:2011hk,Hebecker:2012aw,Palti:2014kza,Blumenhagen:2014gta,Hebecker:2014eua,Grimm:2014vva,Arends:2014qca,Bizet:2014uua,Marchesano:2014mla}.}
 
In this paper we will analyze instead elliptic flux backgrounds and their corresponding effective potentials.
They provide a new class of geometric as well as non-geometric NS string backgrounds. We stress that a full understanding of truly non-geometric solutions is still lacking and quite challenging, since ten dimensional supergravity is of little use and in general one expects significant string corrections. While in some cases a worldsheet approach can be used (see for example~\cite{Becker:2006ks,Anastasopoulos:2009kj}), it would be highly desirable to have an effective description that capture some of the main features of such backgrounds. One can indeed make progress from a supergravity description in four dimension~\cite{Dabholkar:2002sy,Shelton:2005cf,Shelton:2006fd}. 
In this paper we will make use of recent works \cite{Hohm:2013bwa,Hassler:2014sba} to study ten dimensional consistency of such effective description\footnote{Ten-dimensional effective actions for non-geometric fluxes were also considered in~\cite{Andriot:2012an,Andriot:2012wx,Andriot:2011uh,Andriot:2013xca,Blumenhagen:2012nk,Blumenhagen:2013hva}. See also~\cite{Grana:2008yw} for a discussion in the context of generalized geometry.} by using dimensional reduction of double field theory (DFT)~\cite{Siegel:1993th,Hull:2009mi,Hull:2009zb,Hohm:2010jy, Hohm:2010pp}.
We will use the DFT tools to get some controls over nongeometric compactifcations, to derive potentials in particular for elliptic backgrounds and to study their vacuum structure and their inflationary properties. The considered backgrounds are intrinsically non-geometric, since they do not lie on an orbit which is T-dual to any known geometric compactification. We will show that the associated potentials with elliptic fluxes do allow for Minkowski vacua and we provide evidence that they can lead to de Sitter vacua. We will also analyze the finite, elliptic monodromies in more detail. In contrast to the parabolic monodromies, they will lead to an enhanced field range not only for the axionic component of the complex modulus, but also for its imaginary part, which in particular describes the volume of the fibre torus inside the compact space. 

We will then demonstrate that these monodromies can be nicely used for inflation, as they in turn lead to an enhanced field range for the inflaton field. Furthermore they also allow for the stabilization of the $\rho$ and $\tau$ moduli of the fibre torus. This is possible since new terms will appear in the potential~\eqref{eq:pot}, created from the interplay  between different fluxes, which will allow the uplift to Minkowski or de Sitter vacua.
 
We will also study if these elliptic potentials can be used to obtain a viable inflation, meeting some of the phenomenologycal restrictions from recent observations. We will show that we can obtain a potential which is almost quadratic in the directions of the real parts of the K\"ahler or complex structure moduli fields, thus providing a simple embedding of a chaotic inflation potential $V(\phi)=m^2\phi^2$~\cite{Linde:1983gd}. We will then study if one can satisfies slow roll conditions for such potentials, and discuss about the allowed field range for the inflaton. As we will see, slow roll conditions will be possible only if one could tune the fluxes to be very small. However, a mildly enhanced field range for the inflation will be in fact possible due to elliptic monodromies of the backgrounds. 

We stress however that these models, being basically the products of two 3-dimensional flux backgrounds, still have to be regarded as toy models. It is plausible that by considering more general six-dimensional spaces, it is possible to obtain different flux combinations that can naturally solve the slow roll conditions, thus giving more realistic cosmological models.\\

\textbf{Note added}. After this paper appeared online, it was found that the observation of large tensor-to-scalar ratio from the BICEP2 collaboration is due to a previously overlooked dust contribution. This reduces the need of large trans-Planckian field ranges for the inflaton and fit nicely with our elliptic models, which favour mild field range enhancements.

\section{Effective potentials}\label{sec:2}

In this section we will consider a very simple model of T-fold
backgrounds, closely related to asymmetric 
orbifolds~\cite{Condeescu:2013yma}, which illustrates some of the
properties that we discussed before.
Recently, this kind of backgrounds have also been studied from
the point of view of consistent dimensional reduction of double field
theory~\cite{Hassler:2014sba}, a fact that makes the study of our
toy model more tractable.
These spaces are non-geometric generalizations of Nilmanifolds (aka twisted tori), where geometric as well as non-geometric NS fluxes can coexist at the same time.
Furthermore these backgrounds have the property that they are in
general not T-dual to any geometric space.
Starting from the DFT action and 
performing a generalized Scherk-Schwarz reduction, the corresponding background solutions are characterized by the requirement of the vanishing of the generalized DFT Ricci
tensor. For simplicity, we will only consider the case of three compact dimensions, 
in which there are only two distinct solutions of these DFT field
equations, called single elliptic and double elliptic cases. The
double geometries in the internal directions of both cases correspond
to fibrations where the doubled fiber is a four-dimensional torus
$\mathrm{T}^4$ over a doubled circle as a base.
The double elliptic case
is not T-dual to a geometric description and it exhibits $H$-, $f$-
and $Q$-flux at the same time and it is thus a truly non-geometric
space (it cannot be written in terms of a globally well defined
metric, $B$-field and dilaton).
We remark that in order to be able to
consider such spaces, one need a mild violation of the strong
constraint of DFT. Indeed, the Killing vectors do not
satisfy the constraint and can depend both on the coordinates and
the windings \cite{Hassler:2014sba}.

In the effective lower-dimensional theory these solutions lead to Minkowski vacua with vanishing ground state energy. 
After performing the dimensional reduction, the effective action is described by a particular gauged supergravity potential, 
which is a function of the scalar field moduli of the (doubled) fibre torus, and
whose discrete parameters are the $H$-, $f$- and $Q$-fluxes of the DFT action.
The scalar potentials we obtain in this way exhibit several interesting and new features, which make them
also potentially interesting for cosmology and in particular for
inflation. 

Since they come from a consistent Scherk-Schwarz ansatz, the corresponding effective potentials
are positive semidefinite with zero or positive energy minima in the moduli directions. In fact, the vanishing or, respectively, the positivity of the potential energy at the minimum of the potential is due to
the contributions of geometric and non-geometric fluxes that precisely
cancel each other, so that no extra uplifting terms, such as orientifolds and/or anti NS5-branes,
are needed in order to obtain a zero cosmological constant. 
This leads to hope that the interplay among different kind of fluxes
can evade the (in)famous no-go theorems about inflation and de Sitter
vacua in flux compactifications.

We will now present the potentials obtained with the methods
of~\cite{Hassler:2014sba} and study their minima, as well as the
monodromies of the backgrounds. In the following section, we will examine inflationary properties of
such effective potentials.

\subsection{Potentials from reducing double field theory}\label{sec:potential}

Let us briefly review the main properties of the non-geometric backgrounds obtained in~\cite{Condeescu:2013yma,Hohm:2013bwa,Hassler:2014sba}. In general, we are considering three-dimensional spaces with non-vanishing $H$, $f$, $Q$, $R$ fluxes\footnote{Our discussion is formal since there is no clear definition of a space with $R$-flux. In any case we will shortly set $R=0$.}. They are solutions in a doubled space in which are topologically given as fibrations of a doubled torus over a doubled circle. The fibration is elliptic and of finite order (in contrast to parabolic fibrations, as e.g. the torus with constant H-flux and its T-dual versions, such as Nilmanifolds). Different fibrations can be specified by so-called twist matrices (for details see~\cite{Dabholkar:2002sy}), which are certain elements forming a $\mathbb{Z}_N\times \mathbb{Z}_M$  subgroup of the modular group $SL(2,\mathbb{Z})_\tau\times SL(2,\mathbb{Z})_\rho$. For consistency, the monodromies of these 3-dimensional backgrounds can be only of the following finite orders:\footnote{For higher dimensional spaces, also higher orders in $N$ like $N=12$ are possible.}
\begin{equation}
N=2,3,4,6\, .
\end{equation}
Alternatively, these backgrounds can also be described in terms of $H$, $f$, and $Q$ fluxes. In DFT, all these fluxes are combined into a single, totally antisymmetric object (called covariant fluxes~\cite{Aldazabal:2011nj,Grana:2012rr,Geissbuhler:2013uka,Aldazabal:2013sca}):
\begin{equation}
  \mathcal{F}_{ABC}: \; F_{abc} = H_{abc} \, , \quad F^a{}_{bc} = f^a_{bc} \, , \quad F^{ab}{}_c = Q^{ab}_c \, ,\quad F^{abc} = R^{abc} \, .
\end{equation}

A generalized Scherk-Schwarz compactification of DFT~\cite{Aldazabal:2011nj,Geissbuhler:2011mx} gives rise to the scalar potential
\begin{equation}\label{eqn:scalarpotential}
  V(\mathcal{H}^{IJ}) = -\frac{1}{4} \mathcal{F}_I{}^{KL}
    \mathcal{F}_{JKL} \mathcal H^{IJ} +
    \frac{1}{12} \mathcal{F}_{IKM}
    \mathcal{F}_{JLN} \mathcal H^{IJ} \mathcal H^{KL} \mathcal H^{MN}\, ,
\end{equation}
where $\mathcal{H}^{IJ}$ is the generalized metric of a $T^3$. Keeping in mind the fiber structure explained above, we parameterized this torus with five moduli: the complex structure $\tau=\tau_R + i \tau_I$, the K\"ahler parameter $\rho=\rho_R + i \rho_I$ of the $T^2$ fibre and the radius $R$ of the base. This parameterization gives rise to
\begin{equation}
  \mathcal{H}^{IJ} = \begin{pmatrix}
    R^2 & 0 & 0 & 0 & 0 & 0  \\
    0 & \frac{|\rho|^2}{\rho_I \tau_I} & \frac{|\rho|^2 \tau_R}{\rho_I\tau_I} & 0 & \frac{\rho_R \tau_R}{\rho_I \tau_I} & -\frac{\rho_R}{\rho_I\tau_I} \\
    0 & \frac{|\rho|^2 \tau_R}{\rho_I\tau_I} & \frac{|\rho|^2|\tau|^2}{\rho_I\tau_I} & 0 & \frac{\rho_R|\tau|^2}{\rho_R\tau_R} & -\frac{\rho_R \tau_R}{\rho_I\tau_I} \\
    0 & 0 & 0 & \frac{1}{R^2} & 0 & 0 \\
    0 & \frac{\rho_R\tau_R}{\rho_I\tau_I} & \frac{\rho_R|\tau|^2}{\rho_R\tau_R} & 0 & \frac{|\tau|^2}{\rho_I\tau_I} & -\frac{\tau_R}{\rho_I\tau_I} \\
    0 & -\frac{\rho_R}{\rho_I\tau_I} & -\frac{\rho_R\tau_R}{\rho_I\tau_I} & 0 & -\frac{\rho_R}{\rho_I\tau_I} & \frac{1}{\rho_I\tau_I}
  \end{pmatrix}\,.
\end{equation}
In the following we will deal with four flux parameters:
\begin{equation}\label{eqn:solutionfluxesa}
  H_{123} = H\, ,\quad Q^{23}_1 = Q\,,  \quad  f^2_{31} =f_1\,  ,\quad
  f^3_{12} = f_2 \, .
\end{equation}
Then one 
 obtains the following scalar potential  in the moduli $\tau$ and $\rho$ and $R$.\footnote{This potential is given in the string frame. In order to go to the Einstein frame one has to multiply this
potential by an additional factor of $\rho_I^{-1}$.} It reads
\begin{equation}
  \label{eqn:explicitspotential}
    V (\tau,\rho)= \frac{1}{R^2} \left[ \frac{f_1^2 + 2f_1f_2(\tau_R^2 - \tau_I^2) + f_2^2 |\tau|^4}{2 \tau_I^2}
  + \frac{ H^2 + 2HQ(\rho_R^2 - \rho_I^2) + Q^2|\rho|^4}{2 \rho_I^2}
\right]\, ,
\end{equation}
The base radius R is a flat
direction of the potential and in the following we will set it to
1. The potential exhibits a different
structure than the potential in eq.(\ref{eq:pot}), since there are cross-terms  that are proportional to the product of two
different fluxes, namely $f_1f_2$ and $HQ$ respectively. These new terms that arise only for the elliptic backgrounds will be responsible for allowing stable Minkowski or
de Sitter vacua.

In DFT the fluxes cannot be arbitrarily chosen, but they are restricted by the 
strong constraint in the following way:
\begin{equation}\label{strongconstr}
  H_{123} R^{123} = 0\,, \qquad  
  Q^{23}_1 f^1_{23} = 0\,, \qquad
  Q^{31}_2 f^2_{31} = 0\,, \qquad 
  Q^{12}_3 f^3_{12} = 0\, .
\end{equation}
Requiring furthermore a Minkowski vacuum in the effective seven-dimensional theory, the DFT equations of motion plus the constraint~\eqref{strongconstr} give rise to
\begin{equation}\label{eqn:solutionfluxes}
  H_{123} = Q^{23}_1 = H\,, \quad  Q^{31}_2 = Q^{12}_3 = 0\,, \quad
  R^{123} = f^1_{23} = 0  \quad \text{and} \quad f^2_{31} = f^3_{12} = f\,.
\end{equation}
As discussed in details in~\cite{Hassler:2014sba}, there are only three
kind of solutions:
\begin{itemize}
  \item[(i)] $f\neq0$, $H=0$. This is a geometric background, analogous to
    a symmetric $\mathbb{Z}_N$ orbifold~\cite{Condeescu:2013yma},
    constructed as a fibered torus with an elliptic monodromy.
  \item[(ii)] $f=0$, $H\neq0$. This is a non-geometric T-fold,
    obtained by a T-duality\footnote{We remark that there is a global
      obstruction to perform this T-duality (see for
  instance~\cite{Hull:2006qs,Belov:2007qj}).} from the solution (i) and correspond to an
asymmetric $\mathbb{Z}_2$ orbifold. It
 has an elliptic monodromy in the K\"ahler parameter.
  \item[(iii)] $f\neq0$, $H\neq0$. This is a non-geometric background
    that does not belong to any T-dual orbit of a geometric
    space and corresponds to an asymmetric $\mathbb{Z}_N$ orbifold. In
    the double field theory formulation, two coordinate
    charts are patched with diffeomorphisms, B- and
    $\beta$-transformations at the same time \cite{Hassler:2014sba}. While the strong
    constraint for the fluxes is satisfied, we stress that the Killing
    vectors violate this constraint and they have dependence on both
    the coordinates and momenta. These spaces have two elliptic
    monodromies in both complex and K\"ahler structure.
\end{itemize}

For a fibration which corresponds to an order $N$ twist, the fluxes $f_1,f_2,H,Q$ have to be quantized in units of $1/N$. This is summarized in table \ref{tab:quantizedfluxes}.
\begin{table}[t]
  \centering
  \begin{tabular}{|c||c|c|}
    \hline
   $N$&  $f_1,f_2 \bmod 1$ & $\tau^\star$ \\
    \hline
   $6$ &$1/6$ & $(1+\sqrt{3} i)/2$ \\
    $4$&$1/4$ & $i$ \\
   $3$& $1/3$ &  $(1+\sqrt{3} i)/2$ \\
   $2$& $1/2$  & $i$ \\
    \hline
  \end{tabular}\qquad\qquad\qquad
  \begin{tabular}{|c||c|c|}
    \hline
  $N$ & $H,Q \bmod 1$ &  $\rho^\star$ \\
    \hline
 $6$ &   $1/6$ &  $(1+\sqrt{3} i)/2$ \\
 $4$ &   $1/4$ &  $i$ \\
   $3$ & $1/3$ &  $(1+\sqrt{3} i)/2$ \\
  $2$ &  $1/2$ &  $i$ \\
    \hline
  \end{tabular}
  \caption{Quantized values for the fluxes (in flat indices) and the corresponding  parameter of the vacuum generalized vielbein $E^\star_A{}^M$ (see appendix B).}\label{tab:quantizedfluxes}
\end{table}
Now it is straight forward to obtain the scalar potential for the $\mathbb{Z}_4$ case. 
Here we have that $f_1=f_2=H=Q=1/4$. Then the scalar potential (\ref{eqn:explicitspotential}) reads:
\begin{equation}
  \label{eqn:explicitspotentialz4}
    V (\tau,\rho)= \frac{1}{16R^2} \left[ \frac{1 + 2(\tau_R^2 - \tau_I^2) +  |\tau|^4}{2 \tau_I^2}
  + \frac{ 1 + 2(\rho_R^2 - \rho_I^2) + |\rho|^4}{2 \rho_I^2}
\right]\, .
\end{equation}
For  the $\mathbb{Z}_6$ case with $N=6$ monodromy and  with $f_1=f_2=H=Q=1/6$ one gets from eq.(\ref{eqn:explicitspotential}) after a field
redefinition the following scalar potential:
\begin{equation}\label{z6potential}
  V = \frac{2}{3 \rho_I^2} \left[ \rho_I^4 + (1-\rho_R + \rho_R^2)^2 + \rho_I^2(-1 -2\rho_R + 2\rho_R^2) \right]\,.
\end{equation}
As mentioned before, the scalar potentials \eqref{eqn:explicitspotential},  \eqref{eqn:explicitspotentialz4} and 
\eqref{z6potential}
correspond to three-dimensio\-nal compactifications from ten to seven dimensions. In order
to obtain a four-dimensio\-nal effective theory, we can consider the
product of two three-dimensional non-geometric backgrounds. In this
case we have a potential
$V (\tau^{(i)},\rho^{(i)})$ ($i=1,2$) with two complex structure
fields $\tau^{(i)}$ and two K\"ahler fields $\rho^{(i)}$. This would
be simply a sum of the two potentials~\eqref{eqn:explicitspotential}:
\begin{equation}
V (\tau^{(i)},\rho^{(i)})=V^{(1)}+V^{(2)}\, .
\end{equation}
In this case we thus have eight different fluxes
$H^{(i)},f_1^{(i)},f_2^{(i)},Q^{(i)}$. 
However it would be also worth to consider other spaces with more general fluxes, which are not of this direct product structure.

\subsection{Vacuum structure}

We now show that the above potentials exhibit non-trivial stable minima due to the simultaneous presence of different kinds of fluxes. 
Since the potentials contain two independent terms,
i.e. $V=V(\tau)+V(\rho)$, let us focus only  on one term, say $V(\rho)$. The discussion for the other term is completely analogous.

\subsubsection{Minkowski vacua}

Let us first analyze the general potential given in eq. \eqref{eqn:explicitspotential}.
Here we make the choice that the product $HQ>0$. Then the potential eq.(\ref{eqn:explicitspotential}) possesses the following stable Minkowski minimum:
\begin{equation}
\rho_R^{\star}=0\,,\quad\rho_I^{\star}=\sqrt{H\over Q}\,,\qquad V_{\rm min}=0\, .
\end{equation}
For the $\mathbb{Z}_4$ case the potential \eqref{eqn:explicitspotentialz4} has its Minkowski minimum at
\begin{equation}
  \rho_R^{\star} =0\,,\quad \rho_I^{\star} = 1\, ,
  \end{equation}
and
for the $\mathbb{Z}_6$ case the potential \eqref{z6potential} has its Minkowski minimum at
\begin{equation}
  \rho_R^{\star} = \frac{1}{2}\,,\quad \rho_I^{\star} = {\sqrt{3}\over{2}}\, .
 \end{equation}
With respect to this two scalar fields, we have a positive definite mass matrix in both cases.

\subsubsection{De Sitter vacua}

Now we relax the constraint by giving up the
requirement of having a Minkowski vacuum with vanishing potential in
the effective theory. 
Namely we want to assume that $H$ and $Q$ have different relative signs, i.e. $HQ<0$.
In this case one obtains a moduli space of stable de Sitter vacua with the following values for the moduli and the potential (here for the $\mathbb{Z}_4$ case):
\begin{equation}
\label{eqn:rhomindS}
(\rho_R^{\star})^2+(\rho_I^{\star})^2=-{H\over Q}\,,\qquad V_{\rm min}=-{4HQ}>0\, .
\end{equation}
In oder to get a very small cosmological constant, one would therefore
need very tiny values for the fluxes $H$ and $Q$. However we note that the radius $R$ is not a flat direction any more:
\begin{equation*}
   \left.\frac{\partial V}{\partial R}\right|_{\rho^{\star}} =  2 \frac{H Q}{R^3} < 0\,.
 \end{equation*}
A potentially better option is to consider a different class of potentials,
derived in Appendix A. For example, for an $SO(2,2)$ gauging we obtain
the potential:
\begin{equation}
 V(\rho,\tau) = \frac{H^2}{2 \rho_I^2} \left[ 1 + 2(\rho_R^2 - \rho_I^2) + |\rho|^4 \right] +
    \frac{H^2}{\rho_I\tau_I} ( 1 + |\rho|^2 )( 1 + |\tau|^2) +
    \frac{H^2}{2 \tau_I^2} \left[1 + 2(\tau_R^2 - \tau_I^2) +
      |\tau|^4 \right] \, .
\end{equation}
This potential is only valid up to second order in the moduli fields $\tau$ and $\rho$. For higher orders one should also consider the five remaining moduli to a $T^3$. The minimum of the potential is located at
\begin{equation} 
  \rho^{\star} = \tau^{\star} = i \quad \text{with} \quad V_{\rm min} = 4 H^2 \, .
\end{equation}
One can also check that the four moduli $\rho_R, \rho_I, \tau_R,
\tau_I$ have all a positive mass of value $8 H^2$.

\subsection{Monodromies and field range}
\label{monodromies}

\begin{figure}[b]
\centering
\includegraphics[width=0.7\textwidth]{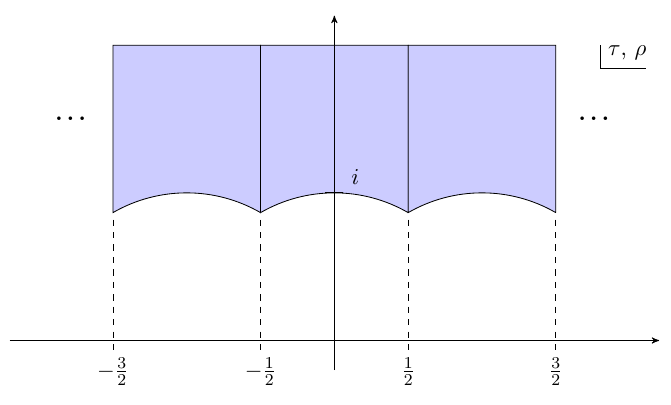}
\caption{By introducing a parabolic monodromy, the field range of $\rho_R$ ($H$-flux) or $\tau_R$ ($f$-flux) is enlarged from $[-1/2,1/2)$ to $(-\infty,\infty)$.}
\label{fig:funddomparabl}
\end{figure}

In this chapter will analyze the  monodromies and the field ranges of the scalar fields in more detail.
The monodromies of the background are always given in terms of some particular group element $\gamma_0$ of $SL(2,\mathbb{Z})$.
Furthermore recall that $SL(2,\mathbb{Z})$ can be generated by the following group elements, denoted by $T$ and $S$, which act on $\rho$ (or $\tau$) in the following way:
\begin{equation}
T:\quad \rho\rightarrow \rho+1\,,\quad S:\quad\rho\rightarrow -{1\over \rho}\, .
\end{equation}
However one can also choose the  two group elements $T$ and $TS$ as generators of the modular group:
\begin{equation}
T:\quad \rho\rightarrow \rho+1\,,\quad TS:\quad\rho\rightarrow -{1\over \rho}+1\, .
\end{equation}
For the original monodromy inflation model of \cite{Silverstein:2008sg,McAllister:2008hb}, the shift symmetry of the potential with respect to the axionic field is broken by a parabolic monodromy which is  induced by a geometric $f$-flux or respectively by its T-dual $H$-flux. Therefore, in this case the monodromy is with respect to the $SL(2,\mathbb{Z})$  group element
$\gamma_0=T$. It follows that due to the non-trivial monodromy the fundamental domain of $SL(2,\mathbb{Z})$ gets unfolded infinitely many times, since the shift is
of infinite order.
This case is depicted in figure \ref{fig:funddomparabl}. As one can see, the a priori finite field range of the axionic scalar field becomes indefinitely large due to the monodromy, i.e. due to the presence of the flux. Furthermore one can couple the closed string background to other probes, like D4-branes or D7-branes, which also feel the monodromy of the background and can be used to produce an inflaton potential.

In this paper we will discuss a slightly different case. Again we start with complex scalar fields, $\tau$ or $\rho$, whose field range is a priori restricted by the requirement that these fields can take only values inside the well-known  fundamental domain of $SL(2,\mathbb{Z})$. Now we will consider non-trivial elliptic monodromies which are
associated to finite order elements $\gamma_0$ of $SL(2,\mathbb{Z})$. Specifically we will take for $\gamma_0=S$ or $\gamma_0=TS$. They
will again unfold the fundamental domain of $SL(2,\mathbb{Z})$, leading to a bigger region in field space, in which the fields can take their allowed values. In this way the original $SL(2,\mathbb{Z})$ symmetry of the theory gets broken by the non-geometric fluxes to a particular subgroup of $SL(2,\mathbb{Z})$, which possesses an enlarged fundamental region. Note that this effect will happen for the real as well as the imaginary parts of $\tau$ or $\rho$. In particular the imaginary fields, corresponding to the torus fibre volume in case of $\rho$, can take all possible positive real numbers, allowing also for sub-stringy volumes of the torus. This will then also allow for an enlarged field range of the inflation field $\phi$, which is a combination of real and imaginary part of $\rho$, as we discuss in the next section of the paper. Also note that we do not need any other probes like D-branes or NS 5-branes for this effect to happen. In fact, the probes that feel the non-trivial monodromy are themselves (massive) closed strings.

\begin{figure}
\centering
\includegraphics[width=0.9\textwidth]{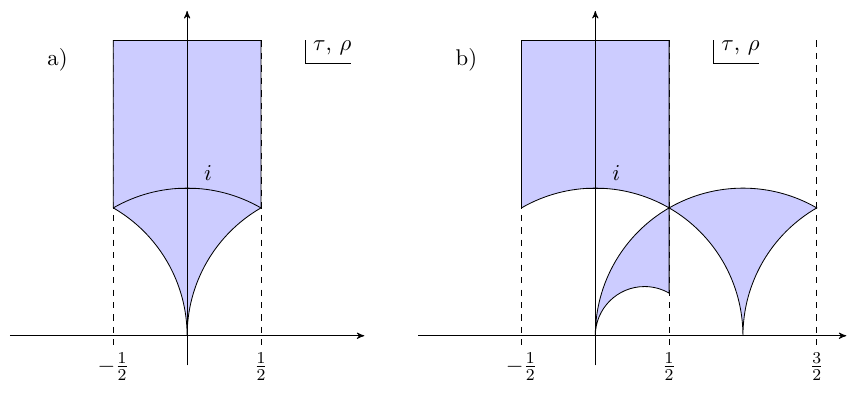}
\caption{Fundamental domain for $SL(2,\mathbb{Z})/\mathbb{Z}_4$ (a) and for $SL(2,\mathbb{Z})/\mathbb{Z}_6$ (b).}
\label{fig:funddomelliptic}
\end{figure}

In order to show the monodromy properties of the corresponding background, we consider $\rho(\mathbb{X})$ as a function of the base coordinate $\mathbb{X}$. For the case $N=4$ the monodromy group elements is $\gamma_0=S$. Then for the potential of eq.(\ref{eqn:explicitspotentialz4}), we explicitly compute the monodromies of $\tau$ and $\rho$ as follows:
\begin{equation}\label{monodromy}
  \tau(\mathbb{X}) = \frac{ \tau(0) \cos(f \mathbb{X}) + \sin(f \mathbb{X}) }{
    - \tau(0) \sin(f \mathbb{X}) + \cos(f \mathbb{X})}\,, \quad
  \rho(\mathbb{X}) = \frac{ \rho(0) \cos(H \mathbb{X}) + \sin(H \mathbb{X}) }{
    - \rho(0) \sin(H \mathbb{X}) + \cos(H \mathbb{X})}\,.
\end{equation}
Here $\tau(0)$ and $\rho(0)$ correspond to the scalar fields $\tau$ and $\rho$ in the scalar potential~\eqref{eqn:explicitspotentialz4}. The order of the monodromy of the solutions is then determined by the quantized values of the fluxes $H=1/4~{\rm mod~1}$ and $f=1/4~{\rm mod~1}$. Plugging the vacuum expectation values $\tau=\rho=i$ into the functions \eqref{monodromy}, one realizes that $\tau(\mathbb{X})$ and $\rho(\mathbb{X})$ themselves become constants and they are thus independent from the value of the base coordinate $\mathbb{X}$. In this case, the geometry is that of a left-right (a)symmetric orbifold. Hence the Minkowski minima of the potential just corresponds to orbifold points in the underlying geometry.
The action of monodromy on the moduli is given by:
\begin{equation}
\rho(\mathbb{X}+1)=-1/\rho(\mathbb{X})\, ,
\end{equation}
and similarly for $\tau(\mathbb{X})$. It follows that the field range of $\rho$ gets enlarged, namely the possible values for $\rho$ are now given by the original fundamental domain ${\cal F}_\rho$ of $SL(2,\mathbb{Z})$ plus its image ${\cal F}_{\tiny \rho\rightarrow -1/\rho}$ under the transformation $\rho\rightarrow -1/\rho$:
\begin{equation}
{\cal F}={\cal F}_\rho\oplus {\cal F}_{\tiny \rho\rightarrow -1/\rho}\, .
\end{equation}
This is depicted in figure \ref{fig:funddomelliptic}a).  We see that in particular the allowed  field range of $\rho_I$ due to the non-trivial monodromy gets enlarged to
\begin{equation}
\rho_I\in [0,\infty ]
\, .
\end{equation}
Therefore also sub-stringy values for the volume of the two-torus are possible.

Next let us discuss the case $N=6$. Here the monodromy is of order six and $\gamma_0=TS$. The functions $\rho(\mathds{X})$ and $\tau(\mathds{X})$ now take the following form:
\begin{align}
\rho(\mathbb{X})&=\frac{ \sqrt{3} \cos(2 \pi H \mathbb{X}) \rho(0) + \sin(2 \pi H \mathbb{X})(\rho(0)-2)}{ \sin(2 \pi H \mathbb{X})(2\rho(0)-1)  + \sqrt{3} \cos(2 \pi H \mathbb{X})} \quad\text{and}\\
  \tau(\mathbb{X})&=\frac{ \sqrt{3} \cos(2 \pi f \mathbb{X}) \tau(0) + \sin(2 \pi f \mathbb{X})(\tau(0)-2)}{ \sin(2 \pi f \mathbb{X})(2\tau(0)-1) + \sqrt{3} \cos(2 \pi f \mathbb{X})}\,.
\end{align}
Now, we get that 
\begin{equation}
  \rho(\mathbb{X}+1)=-{1\over\rho(\mathbb{X})}+1\, \quad{\rm and} \quad\rho(\mathbb{X}+2)={1\over 1 -\rho(\mathbb{X})}
\end{equation}
and similarly for $\tau(\mathbb{X})$. The allowed values for $\rho$ are now given by the sum of the three fundamental domains
\begin{equation}
{\cal F}={\cal F}_\rho\oplus {\cal F}_{\tiny \rho\rightarrow -{1\over\rho}+1}\oplus{\cal F}_{\tiny\rho\rightarrow {1\over1-\rho}} \, .
\end{equation}
This fundamental domain is depicted in \ref{fig:funddomelliptic}b).
We see that the field ranges both for $\rho_R$ as well as for $\rho_I$ are enlarged due to the nontrivial monodromy:
\begin{equation}
  \rho_R\in \left[{1\over 2},{3\over 2} \right]\, ,\quad\rho_I\in [0,\infty ]\, .
\end{equation}

\section{Non-geometric inflation}

In this section we will apply some of the properties of nongeometric
backgrounds discussed above to study possible embeddings of
inflationary potentials.
Before we can analyze the inflationary properties of the non-geometric flux scalar potential we first have to go to a scalar field basis in which the inflation field has canonical
kinetic energy. This can be done by knowing the K\"ahler potential of the complex fields $\rho$ and $\tau$ in supergravity, which takes the following form
\begin{equation}
\mathcal{K}(\tau,\rho)=-\log \big[ i(\tau-\bar\tau)\big]-\log\big[i(\rho-\bar\rho)\big]\, .
\end{equation}
It follows that the kinetic energy Lagrangian for the scalar fields takes the following form\footnote{This agrees with the kinetic energy computed directly in DFT.}:
\begin{equation}
{\cal L}_{\rm kin}={1\over 4\tau_I^2}\bigg[(\partial \tau_I)^2+(\partial \tau_R)^2\bigg]+{1\over 4\rho_I^2}\bigg[(\partial \rho_R)^2+(\partial \rho_I)^2\bigg]\, .
\end{equation}
In the following we will simplify the discussion by considering only
the $\rho$-dependence of the potential. The real part of $\rho$ will basically
play the role of the inflaton field $\phi$, whose canonical normalization is:
\begin{equation}\label{inflatonfield}
\phi={\rho_R\over 2\rho_I}\, .
\end{equation}
Therefore we see that the inflaton is actually a particular combination of $\rho_R$ and $\rho_I$ and
hence the range of the inflation field $\phi$ is determined by the range of $\rho_R$ as well as by the range of $\rho_I^{-1}$.
It can be large, provided $\rho_R$ can take large values, or alternatively, if the volume $\rho_I$ of the two-torus can be smaller than the string scale. We will come back to this
issue at the end of this section.
Our model for the inflaton potential then takes the following form:
\begin{equation}\label{inflatonpotential}
V(\phi, \rho_I)=V_0(\rho_I) +m^2(\rho_I)\, \phi^2+\lambda (\rho_I)\,\phi^4\, ,
\end{equation}
where we defined:
\begin{equation}
V_0(\rho_I)= \frac{ H^2  - 2HQ\rho_I^2 + Q^2\rho_I^4}{2 \rho_I^2}\, ,
\quad
m^2(\rho_I)=   4HQ + 4Q^2\rho_I^2\, , \quad
\lambda (\rho_I)= 8Q^2\rho_I^2\, .
\end{equation}
For now we treat the fluxes $H$ and $Q$ as free
parameters. 
For a given model with elliptic monodromy of order $N$ we expect that
\begin{equation}
H\, , Q\sim {1\over N}\, .
\end{equation}
We will discuss shortly more constraints on the fluxes. 
Note that we also need to reintroduce appropriate mass dimensions.
In the convention of dimensionful coordinates, the volume $\rho_I$ is
dimensionless, while $\rho_R$ and the fluxes have mass dimension
one. Also, we need to switch to the Einstein frame, which would amount
to multiply the potential by one inverse power of the volume, i.e. by
$\rho_I^{-1}$ (we are setting all other dimensions to one).

It is now easy to see that the potential is minimized when $V_0 = 0$,
which fixes the value of $\rho_I$ to a given value
$\rho_I^{\star}$. Plugging back this value in the potential then one
gets the parameters:
\begin{equation}
m^2=  4 H' Q'  \left(\frac{1}{\rho_{I}^{\star}}+\rho_{I}^{\star}\right)M_s^{2}\, , \qquad
\lambda =8 H'Q' \rho_{I}^{\star}\, ,
\end{equation}
where we defined dimensionless fluxes $H'$ and $Q'$ with $H',Q'\sim{1\over N}$.

Now we can study the  inflationary behavior of the potential  with respect to $\phi$. For this purpose we have to write the potential not as a function of the string mass $M_s$ but rather
of the effective 4-dimensional Planck mass $M_P$, given by
\begin{equation}
M_P^2={1\over g_s^2}M_s^2\rho_I^{\star} \, .
\end{equation}
Performing this change of masses, we get the following result for the
ratio between quartic and quadratic coupling:
\begin{equation}
g_s^2M_P^2\frac{\lambda}{m^2} =
\frac{2\,(\rho_{I}^{\star})^3}{1+(\rho_{I}^{\star})^2} \, .
\end{equation}
We are interested in small value of $\rho_{I}^{\star}$, for which this
ratio would be given by $(\rho_{I}^{\star})^3$.
 It is thus controlled by the volume of the torus fibre.
Hence, we
learn that by requiring the quartic coupling
$\lambda$ to be much smaller then the mass $m^2$, we force the volume of the fibre torus to be smaller
then the string scale! As one can see from eq.(\ref{inflatonfield})
 this is precisely the condition
which can enlarge the inflaton field range. As we discussed in
section~\ref{monodromies}, such sub-stringy volumes are in principle allowed by the
elliptic monodromies and the breaking of the $SL(2,\mathbb{Z})$ symmetry. In our simple $\mathbb{Z}_4$ or $\mathbb{Z}_6$
models one can get only a mild enlargement, since $\rho_I^{\star}$ corresponds to the orbifold limit
which is fixed by the order of the monodromy. However, it is likely
that one can construct models with higher order monodromy which can
enlarge the inflaton range up to the phenomenologically relevant value,
and contribute to suppress the quartic coupling in our model.
One can also make a short comment on higher order $\phi^n$ ($n>4$) terms in the potential. One can perhaps expect that the order
$n$ term gets a suppression which is of some power of $\rho_I^{\star}$. In this case, since we are working at small sub-stringy volumes, these higher terms would
not destroy the inflationary behavior of the scalar potential.

This
raise the question of whether it makes sense to consider a
volume smaller than the string length, since this would imply a string
scale higher then then than Planck mass. To avoid this, one could try
to use for example the $\tau$ modulus for inflation, for which one can
apply the very same mechanism discussed above. In this way one can get
a trans-Planckian excursion but one would need to explore more general
models to get a small quartic coupling, and this could be very probably
achieved in a model with more fluxes.
 This investigation would clarify if
the difficulties in obtaining trans-Planckian field excursions in
string theory can be alleviated by using nongeometric backgrounds and
if it is possible to do so in a regime where one can trust the
effective description.

It would also be interesting to try to see if different
monodromies (which perhaps do not admit a fixed orbifold point) could
provide more general mechanisms. One could consider backgrounds with hyperbolic (or sporadic) monodromies. Potentials for this case have appeared for example in~\cite{Schulgin:2008fv}, and this seems a very promising line of investigation. It would also be interesting to study in more generality actions based on coset models with an $SL(2,\mathbb{Z})$ symmetry in the context of inflation.
We also remark that it would be natural to consider models based on hyperbolic Riemann surfaces. 
For example, for a genus 2 surface the modular group would be $SO(2,3,\mathbb{Z})$ and one would need to consider three parameters $(\tau, \rho, A)$~\cite{Mayr:1995rx,Dijkgraaf:1996it}. One might speculate that this kind of models are relevant for de Sitter compactifications (see for example~\cite{Saltman:2004jh}). \\

At the end of this section let's make a phenomenological fit of the parameters of the potential, which will lead to a viable slow roll inflationary scenario.
This requires that the mass parameter $m$  and, as we discussed before, the quartic coupling are sufficiently small, which can be expressed in terms of the well known conditions
of the inflationary slow roll parameters $\epsilon$ and $\eta$ (see appendix C for the definition of these parameters).

For example, we assume about 60 e-foldings for inflation and 
we fit the recent BICEP2 data \cite{Ade:2014xna}:
\begin{equation}
n_s\sim 0.967\, ,\quad r\sim 0.133\, ,
\end{equation}
which we take to define an upper bound on pur parameters. We then choose the following values for $V_0$ and $m$:
 \begin{equation}
m\simeq 6\times 10^{-6}M_P\, ,\quad V_0^{1/4}\simeq10^{-2}M_P \,\,\Rightarrow \,\,\phi\simeq 15 M_P \, .
\end{equation}
As well known, the relatively large value $r$ of the tensor fluctuations requires a very large trans-Planckian field value for the inflaton field.
For the quartic term to be smaller than the mass term,  one would need to choose $\lambda\leq 10^{-13}$. 
We note that we can fit these number by the following choice of fluxes:
\begin{equation} H'\simeq Q'\simeq 10^{-5}\,,\quad     \rho_I^{\star}    \leq  10^{-2}\, .
 \end{equation}
So we see that in order to fit experimental data, our simple toy model would require the fluxes to be very tiny, or in other words the order of the monodromy must be very large:
$N\simeq 10^5$.
If this is possible in a more realistic model still has to be seen. Furthermore, as discussed already, for the necessary suppression
of the quartic coupling $\lambda$,
 one needs that
the volume of the fibre torus is smaller than the string length$^2$. At the same time the small volume $\rho_I$ also enhances the field range of $\phi$  and hence a large field range for the inflaton field seems possible within the considered framework. 

\section{Conclusion}

In this note we constructed explicit toy models of nongeometric string
compactifications in order to understand if new terms such as nongeometric
fluxes can be used for phenomenologically relevant models in
cosmology. In particular, we focused on the existence of de Sitter vacua
and the possibility of a trans-Planckian field excursion for an
inflaton field. The latter possibility is crucial in view
of recent advances in observational cosmology and in particular the
possibility of a detection of large tensor-to-scalar ratio in the near future. As it is well appreciated, this result implies a
trans-Planckian field excursion for the inflaton field, which
exacerbate UV sensibility of inflation and calls for a treatment in a
UV complete theory of gravity, for which string theory is our best
candidate. However, obtaining such a large field excursion in string
theory is known to be
problematic~\cite{Banks:2003sx,ArkaniHamed:2006dz,Baumann:2006cd}, and most string
models give indeed prediction for a small tensor perturbation (see
for example~\cite{Burgess:2013sla,Baumann:2014nda} for recent
overviews).

In view of this situation, we believe that it is crucial to try to
explore different corners of string theory compactifications, in order
to see if these problems are generic or tied to a specific limit. In
this respect, one promising approach is to explore nongeometric
compactifications. These are backgrounds that do not admit an
interpretation in terms of conventional geometric notions, but
nonetheless their existence is well understood in string theory (a
simple example are for instance Landau-Ginzburg models or asymmetric
orbifolds). In this context, it is possible to obtain Minkowski vacua
with moduli stabilized by fluxes (see for
example~\cite{Becker:2006ks,Anastasopoulos:2009kj}), which is
notoriously difficult in type II supegravities. While very promising, a systematic analysis of nongeometric
backgrounds is quite challenging because in general one should abandon
a large volume limit and stringy effects can become important.

In this note, we took a pragmatic approach and we used recent
developments in the double field theory
formalism to take a grip on
possible effective potentials arising from nongeometric
compactifications and study their vacuum structure, as well as their
inflationary properties. While our results would require computing
string corrections to be put on a firm ground, the perspectives that
are emerging are quite interesting. We find that the presence of
nongeometric fluxes can naturally lead to Minkowski minima and more
efficient moduli stabilization. We also provide some evidence for the
existence of de Sitter minima in this context. It would be very
interesting to study more general examples and to perform a general
analysis of strong energy condition violation (which is required in
order to have de Sitter solutions) by nongeometric sources along the
lines of~\cite{Hertzberg:2007wc}.

Another virtue of nongeometric background is the presence of
monodromies for the moduli. We studied if this can lead to an
enlargement of field ranges for a candidate inflaton field. We find
that in our simple toy model, which consist in a one dimensional base
and a fibered two torus, elliptic monodromies can indeed extend the
field range by an amount that depends on the order of the monodromy at
the orbifold point. Whether at a generic point of moduli space or for
more realistic higher dimensional compactifications this field range
can take larger values is an interesting problem for future
investigation.

We finally analyzed a simple model with Minkowski vacuum and we tried
to identify the inflaton with a combination of real and imaginary parts of the K\"ahler
modulus. The potential for the canonically normalized inflaton field $\phi$
has a simple structure $V(\phi) = m^2\phi^2 + \lambda \phi^4$ which
for small $\lambda$ is in good agreement with the simple chaotic
inflation model~\cite{Linde:1983gd}. The problem of getting a light field is
related to the problem of getting a large monodromy. It is promising to see that this is indeed possible for sub-stringy values of the volume modulus, which are
possible due to the elliptic monodromy. Whether this can eventually realized in a full fledged realistic mode still has to be further explored.

Our results seem to suggest that nongeometric backgrounds are good
candidates to solve, or alleviate, some of the problem of string
phenomenology based on geometric flux compactifications, and provide a
first step toward understanding constraints on effective descriptions
such as those
in~\cite{Shelton:2005cf,Shelton:2006fd,Blaback:2013ht}. We believe however
that much technological progress is still required before general
nongeometric compactifications can be studied in a rigorous way.

In this arduos journey, it comes with a great pleasure
that current and future experimental results in observational
cosmology could provide a guide to discriminate between different
models. For example, it would be interesting to ask if phenomenologically relevant vacua
can arise in a corner of the string landscape compatible with
perturbative calculability.

\subsection*{Acknowledgments}

This work was partially supported by the ERC
Advanced Grant ``Strings and Gravity'' (Grant.No. 32004) and by the DFG cluster of excellence ``Origin and Structure of the Universe''.

\appendix
\section{Solutions compatible with all constraints}
\label{app:fluxclassification}
The following table list all configurations for 3 internal directions which are compatible with the quadratic constraint in double field theory. In general there are two parameters $a$ and $b$ which both have to be positive. All other flux backgrounds which fulfill the quadratic constraint are T-dual to the backgrounds in this table.
\begin{center}
\begin{tabular}{|c||cccc|cccc|c|c|c|}
  \hline
  ID & $H_{123}$ & $Q_1^{23}$ & $Q_2^{31}$ & $Q_3^{12}$ & $R^{123}$ & $f_{23}^1$ & $f_{31}^2$ & $f_{12}^3$ & SC & EOM & $V_0$ \\
  \hline
  1 & $a$ & $a$ & $a$  & $a$  & $b$ & $b$ & $b$  & $b$  & $b=0$ & OK & $-4 (a+b)^2$ \\
  2 & $a$ & $a$ & $a$  & $-a$ & $b$ & $b$ & $b$  & $-b$ & $b=0$ & $b=a$, $b=-a$ & $-4a$, $12a$ \\
  3 & $a$ & $a$ & $-a$ & $-a$ & $b$ & $b$ & $-b$ & $-b$ & $b=0$ & OK & $ 4(a-b)^2$ \\
  \hline
  4 & $a$ & $a$ & $a$  & $0$  & $0$ & $0$ & $0$  & $-b$ & OK    & $b=a$, $b=-a$ & $-a^2$ \\
  5 & $a$ & $a$ & $-a$ & $0$  & $0$ & $0$ & $0$  & $-b$ & OK    & - & - \\
  \hline
  6 & $a$ & $a$ & $0$ & $0$  & $0$ & $0$ & $b$  & $b$  & OK    & OK  & 0 \\
  7 & $a$ & $a$ & $0$ & $0$  & $0$ & $0$ & $b$  & $-b$ & OK    & $b=0$  & 0 \\
  8 & $a$ & $a$ & $0$ & $0$  & $0$ & $0$ & $0$  & $b$  & OK    & $b=0$  & 0 \\
  9 & $a$ & $-a$& $0$ & $0$  & $0$ & $0$ & $b$  & $-b$ & OK    & -  & - \\
  10& $a$ & $-a$& $0$ & $0$  & $0$ & $0$ & $0$  &  $b$ & OK    & -  & - \\
  \hline
  11& $a$ & $0$& $0$ & $0$   & $0$ & $0$ & $0$  &  $b$ & OK    & -  & - \\
  \hline
\end{tabular}
\end{center}
Assuming the strong constraint in DFT to hold, only the cases 1, 3 and 6 survive:
\begin{itemize}
  \item Case 1 is a $S^3$ with $H$-flux and has SO(4) as gauge group. The scalar potential of this background has an AdS minimum.
  \item Case 3 has SO(2,2) as gauge group and a dS minimum in its potential.
  \item Case 6 corresponds to the single/double elliptic cases discussed in this paper.
\end{itemize}

\section{Generalized vielbein}
In order to evaluate \eqref{eqn:scalarpotential} in section \ref{sec:potential}, one has to connect the covariant fluxes in flat indices with the one in curved ones. This is done by the so called vacuum generalized vielbein $E^\star_A{}^M$ which is given in term of $\rho^\star$ and $\tau^\star$ as
\begin{equation}
  E^\star_A{}^M = \begin{pmatrix}
    1 & 0 & 0 & 0 & 0 & 0 \\
    0 & \sqrt{\frac{\rho_I^\star}{\tau_I^\star}} & \sqrt{\frac{\rho_I^\star}{\tau_I^\star}}\tau_R^\star & 0 & 0 & 0 \\
    0 & 0 & \sqrt{\frac{\rho_I^\star}{\tau_I^\star}} \tau_I^\star & 0 & 0 & 0 \\
    0 & 0 & 0 & 1 & 0 & 0 \\
    0 & 0 & \sqrt{\frac{\tau_I^\star}{\rho_I^\star}}\rho_R^\star & 0 & \sqrt{\frac{\tau_I^\star}{\rho_I^\star}} & 0 \\
    0 & - \frac{\rho_R^\star}{\sqrt{\tau_I^\star\rho_I^\star}} & - \frac{\rho_R^\star \tau_R^\star}{\sqrt{\rho_I^\star\tau_I^\star}} & 0 & -\frac{\tau_R^\star}{\sqrt{\rho_I^\star\tau_I^\star}} & \frac{1}{\sqrt{\rho_I^\star\tau_I^\star}}
  \end{pmatrix}\,.
\end{equation}
It acts on the covariant fluxes with flat indices as
\begin{equation}
  \mathcal{F}^{IJK} = E^\star_A{}^I E^\star_B{}^J E^\star_C{}^K \mathcal{F}_{ABC}\,.
\end{equation}
Informations about the correct vacuum generalized vielbein can not be obtained from DFT since all $O(D,D)$ valued vacuum vielbeins give rise to the same results. Nevertheless one can use additional informations from string theory. There, the global $O(D,D)$ symmetry is broken to $O(D,D,\mathbb{Z})$ parameterizing T-duality. With these extra informations, there is a vacuum generalized vielbein with gives rise to a monodromy in $O(3,3,\mathbb{Z})$ if and only if the fluxes are quantized as listed in table \ref{tab:quantizedfluxes} in section \ref{sec:potential}. Furthermore, this table shows the parameters $\tau^\star$ and $\rho^\star$ of the corresponding vacuum generalized vielbeins.

\section{Useful relations}

We recall a number of relations which are useful in relating effective potentials obtained from string compactifications to inflaton potentials. We define the string length as $l_s = 2\pi\sqrt{\alpha'}$ and the corresponding string mass as $M_s=l_s^{-1}$. The four dimensional reduced Planck mass ($M_P \sim 2.4 \times 10^8 Gev/c^2$) is given by
\begin{equation}
  M_P^2 = \frac{4\pi \mathcal{V}}{g_s^2 l_s^2} \, ,
\end{equation}
where we suppose that the compact manifold has a characteristic length $L$ and we define the dimensionless volume as $\mathcal{V} = L^6/l_s^6$. The relation between string and Planck masses is then
\begin{equation}
  \frac{M_s}{M_P} = \frac{g_s}{\sqrt{4\pi \mathcal{V}}} \, .
\end{equation}
We also recall the standard formulae for slow roll parameters from an inflaton potential $V(\phi)$:
\begin{equation}
  \epsilon  = \frac{M_P^2}{2}\left(\frac{\partial_{\phi}V }{V}\right)^2
  \, ,\qquad \eta = M_P^2 \left(\frac{\partial^2_{\phi} V}{V}\right) \,
  ,
\end{equation}
from which we can get the spectral index and the tensor-to-scalar ratio:
\begin{equation}
  n_s = 1 - 6 \epsilon+2\eta \, ,\quad r = 16 \epsilon \, .
\end{equation}


 \bibliographystyle{utphys}
\providecommand{\href}[2]{#2}\begingroup\raggedright\endgroup

\end{document}